\documentclass[preprint,preprintnumbers,amsmath,amssymb]{revtex4}

\usepackage{graphicx}% Include figure files
\usepackage{dcolumn}% Align table columns on decimal point
\usepackage{bm}% bold math

\begin{document}

%\preprint{APS/123-QED}

\title{\bf Local Structure and Macroscopic Properties in PbMg$_{\rm 1/3}$Nb$_{\rm 2/3}$O$_3$-PbTiO$_3$ and  PbZn$_{\rm 1/3}$Nb$_{\rm 2/3}$O$_3$-PbTiO$_3$  solid solutions}

\author{ Ilya Grinberg  and Andrew M. Rappe}

\address{The Makineni Theoretical Laboratories, Department of Chemistry, University of Pennsylvania, Philadelphia, PA
                19104-6323}

\date{\today}% It is always \today, today,

\begin{abstract}

We have examined the local structure of PbMg$_{\rm 1/3}$Nb$_{\rm 2/3}$O$_3$-PbTiO$_3$ (PMN-PT) and PbZn$_{\rm 1/3}$Nb$_{\rm 2/3}$O$_3$-PbTiO$_3$ (PZN-PT) solid solutions using density functional theory.  We find that the directions and magnitudes of cation displacement can be explained by an interplay of cation-oxygen bonding, electrostatic dipole-dipole interactions and short-range direct and through oxygen Pb-B-cation repulsive interactions.  We find that the Zn ions off-center in the PZN-PT system, which also enables larger Pb and Nb/Ti displacements.  
The off-centering behavior of Zn lessens Pb-B-cation  repulsion, leading to a relaxor to ferroelectric and a rhombohedral to tetragonal phase transition at low PbTiO$_3$ content in the PZN-PT system. We also show that a simple quadratic relationship exists between Pb and B-cation displacements and the temperature maximum of dielectric  constant, thus linking the enhanced displacements in PZN-PT systems with the higher transition temperatures.
\end{abstract}

\maketitle

Heterovalent ferroelectric perovskite solid solutions have been intensively studied~\cite{Ye02p35,Noheda02p27,Ramer98p31,Gehring01p277601,Park97p1804} as they exhibit a range of interesting structural and dielectric properties which make them useful in device applications.   Relating the macroscopic properties of these materials to the microscopic properties of the constituent atoms is an important scientific goal and is vital for the rational design of new materials with improved properties. 
The technologically important~\cite{Park97p1804} PbMg$_{\rm 1/3}$Nb$_{\rm 2/3}$P$_3$-PbTiO$_3$ (PMN-PT) and  PbZn$_{\rm 1/3}$Nb$_{\rm 2/3}$P$_3$-PbTiO$_3$ (PZN-PT) solid solutions are good subjects for an investigation of structure-property correlations due to the availability of a wide range of experimental data and because the small difference between Mg and Zn ionic sizes~\cite{Shannon70p1046} cannot explain the differences  in the characteristics of the two solid solutions such as a higher $T_c$ and dielectric constant observed for PZN~\cite{Landolt}.

Ab-initio Density Functional Theory (DFT) calculations were
performed for the (1-$x$)PMN--$x$PT and (1-$x$)PZN--$x$PT systems using $2 \times 2 \times 2$ 40-atom or $3 \times 2 \times 2$ 60-atom supercells with periodic boundary
conditions at experimental volume~\cite{Landolt}.   The energy of the system was evaluated using a local density
approximation exchange-correlation functional~\cite{Perdew81p5048} and
was minimized with respect to the atomic coordinates by a quasi-Newton
method with no symmetry imposed.  A $2
\times 2\times 2$ $k$-point sampling of the Brillouin zone was used.  The
calculations were done with designed non-local norm conserving pseudopotentials~\cite{Ramer99p12471}\footnote[1]{Ionized +2 reference configurations were used for Pb, Mg, and Zn, ionized +4 and +5 were used for Ti and Nb and a neutral reference configuration was used for O.  Energy cutoff of 50~Ry were used for all elements except for Zn for which 56~Ry was used.  Real space cutoffs were 2.0, 1.83, 1.95, 1.70, 1.75 and 1.5 for Pb, Mg, Zn, Ti, Nb and O respectively.}
We study four 
compositions with $x = 0$, 0.25, 0.625 and 1.0
  For pure PMN  and PZN we used a 60-atom supercell with  B-cations arranged according to the random-site model~\cite{Davies00p159}.  
 A B-cation arrangement consistent with the random-site model as well as two disordered B-cation arrangements were used to model  the 0.25 composition and three disordered B-cation configurations were used for the 0.625 composition.

The excellent agreement between the pair distribution functions (PDF) obtained from our relaxed 60-atom structure for PMN and by neutron-scattering experiments~\cite{Egami98p1} (Figure 1) and similar  agreement between DFT and experimental PDFs in previous studies of PbSc$_{\rm 2/3}$W$_{\rm 1/3}$O$_3$-PbTiO$_3$ (PSW-PT), PbSc$_{\rm 2/3}$W$_{\rm 1/3}$O$_3$-PbZrO$_3$ (PSW-PZ)~\cite{Juhas04p214101} and PbZr$_{1-x}$Ti$_x$O$_3$ (PZT)~\cite{Grinberg02p909,Grinberg04p144118} solid solutions indicate  that the size of our supercells is sufficient for capturing the local structure of these materials.

We find that at all compositions PZN-PT and PMN-PT local structure is governed by three main interactions.  First, bonding interactions with oxygen favor large Pb and B-cation displacements (Table I) and creation of short Pb-O and B-O bonds.   Second, short-range repulsive Pb-B-cation interactions create anisotropy in the potential energy surfaces felt by the Pb cations, leading to a variation in the preferred local displacement direction for the Pb atoms (Table I, Figure 2).  Third, electrostatic interactions favor dipole alignment and tend to minimize the scatter in displacement direction, eliminating displacements opposite to the overall polarization of the supercell (Figure 2).  Breathing motion by the B-O$_6$ octahedra (to achieve preferred octahedral volume) and small octahedral rotations (0-5$^\circ$) are also present.

	Examination of the data in Table I shows that for both solid solutions Pb displacement magnitudes predicted by DFT  are unchanged by the addition of PT, whereas the average magnitude of the B-cation displacements steadily increases with PT content.    The cation displacements are larger in PZN-PT than in PMN-PT, with especially large difference for the B$^{\rm 2+}$ cations.  While Mg ions are only slightly displaced, the Zn ions exhibit significant distortions even in pure PZN and Ti-like 0.27~\AA $ $ distortions in Ti-rich PZN-PT. As will be explained later, the large Zn distortions also account for the larger magnitude of Pb, Nb and Ti distortions in the PZN-PT system.

	   The contrast between the absence of Mg off-centering and the large Zn displacements is due to the different electronic structure of the two ions.  While Mg is a simple metal and Mg$^{2+}$-O bonding is essentially ionic, the imperfect screening of the nuclear charge by the $d$-electrons  and the presence of low-lying $p$ orbitals make Zn more polarizable and enable covalent bonding with oxygen atoms ~\cite{Cohen92p136,Halilov03p3443}.

As in other Pb containing ferroelectric perovskite solutions, Pb displacements are the determining factor for the average structure of our solid solutions~\cite{Grinberg02p909,Grinberg04p144118,Dmowski00p229}.  
Inspection of Figure 2 shows that the location of Mg atoms is the primary influence on the direction of Pb distortions.  In all cases Pb atoms avoid cube faces with three Nb atoms and one Mg atom and move toward cube faces with two Mg and two Nb atoms.
 The avoidance of Nb-rich faces is due to the presence of oxygen atoms with two Nb neighbors. Such oxygen atoms have higher B-O bond order than the oxygen atoms in PbTiO$_3$ due to the higher valence of Nb.  Since the total bond order of cation-oxygen bonds is conserved at 2, this prohibits the formation of short Pb-O bonds of high bond order and creates a strong effective Pb-Nb repulsive interaction.  On the other hand, the oxygen atoms  with Mg and Nb neighbors  have a lower total B-O bond order than the oxygen atoms in PbTiO$_3$,  necessitating a formation of more short Pb-O bonds to compensate  for B-O bond depletion~\cite{Cockayne99pR12542,Grinberg04p144118} and leading to a weaker effective Pb-Mg repulsion.

The larger Pb and B-cation displacements in the PZN-PT system are due to the coupling between Zn and Pb and Nb off-center displacements through the Pb-B-cation repulsive interactions.   The ionic size of Mg and Zn atoms is essentially the same~\cite{Shannon70p1046}, which leads to very similar shortest allowed Pb-Mg and Pb-Zn distances.   The greater ability of Zn atoms to off-center allows Pb atoms in PZN-PT solution to preserve the required Pb-B-cation distances for larger Pb displacements than in the PMN-PT system.  The larger Pb displacements in turn cause larger Nb and Ti distortions to prevent high Pb-B-cation repulsion at short Pb-Nb and Pb-Ti distances.

The electrostatic dipole-dipole interactions favor alignment of cation displacements and prevent cation displacements in the ($\bar{1}$00) direction in direct opposition to the overall polarization along [100] in Figure 2. The local direction of Pb displacement is determined by a balance between the need to align with other cation displacements and the need to minimize direct and through-oxygen Pb-B-cation repulsion~\cite{Grinberg02p909,Grinberg04p144118}.

At high Ti content, collinear [100] Pb atom distortions create short Pb-O bonds, maximize dipole alignment and minimize local Pb-B-cation repulsion by avoiding the Ti cations located along the [111] direction.  As Ti content decreases, the Pb local environments become anisotropic due to the presence of the overbonded and underbonded oxygen atoms which block Pb distortions toward some [100] perovskite cube faces. This leads to a transition to a rhombohedral phase, where Pb cations move in a variety of low symmetry directions around the overall [111] polarization direction toward the available low-repulsion cube faces.  With the onset of random-site B-cation ordering, the  high anisotropy created by a large amount of faces blocked by the presence of overbonded oxygen atoms forbids the formation of a long-range ferroelectric state even in the [111] direction and leads to a relaxor phase.  A similar explanation, albeit focused on the B-cation size difference, for the relaxor to ferroelectric transition in the ordered PMN-PSN system,  was proposed by Farber and Davies~\cite{Farber03p1861}.

While exact Pb-environment population analysis in PMN-PT and PZN-PT is currently not possible due to the complex and unknown nature of the short-range B-cation ordering in these systems~\cite{Burton02p1359,Cockayne99pR12542}, the above framework coupled with the differences in Zn and Mg distortion behavior explains why in the PZN-PT system the tetragonal phase is present at lower PT content (10$\%$) than in the PMN-PT system (35$\%$), as well as the recent discovery of ferroelectric phases even in pure PZN itself~\cite{Bing04preprint}.   A large Zn distortion softens Pb-Zn repulsion, making perovskite faces with Zn atoms more friendly toward a Pb distortion than the corresponding perovskite faces with Mg atoms in PMN.  This increases the population of the  local environments favorable to tetragonal distortion, moving the relaxor to ferroelectric transition and the morphotropic phase boundary (MPB)  to lower Ti concentrations.

	We find that there is also a strong relationship between the low-temperature cation displacements as found by our DFT calculations and the experimentally obtained  temperature position of the dielectric constant maximum $T_{\epsilon,\rm max}$. 
A qualitative relationship between average lattice displacement and ferroelectric to paraelectric phase transition temperature $T_c$, which is closely linked to $T_{\epsilon,\rm max}$, was found by Abrahams, Kurtz and Jamieson (AKJ) in 1968~\cite{Abrahams68p551}.  However, examination of the original AKJ paper shows that the largest discrepancies between predicted and experimental $T_c$ are for the Pb containing perovskites PbFe$_{1/2}$Nb$_{1/2}$O$_3$ and PT.  We find that $T_{\epsilon,\rm max}$ at 1~MHz for PMN-PT, PZN-PT, PZT, PbSc$_{\rm 1/2}$Nb$_{\rm 1/2}$O$_3$ (PSN), PSW-PT and PSW-PZ solid solutions is predicted by 

\begin{eqnarray}
 T_{\epsilon,\rm max} = a d_{\rm Pb}^2 + b d_{\rm B'}^2 f_{\rm B'},
\end{eqnarray}      

\noindent where $d_{\rm Pb}$ is the average magnitude of the Pb distortions, $d_{\rm 
B'}$ is the average magnitude of the distortions, $f_{\rm 
B'}$ is the fraction of the ferroelectrically active B-cations in solution and   $a$ and $b$ are 
constants (1739 and 5961 respectively) in units of K/\AA$^2$.

The dependence of $T_{\epsilon,\rm max}$ on local structure presented in Equation 1 can be further simplified by transforming cation and oxygen displacements ${\mathbf u}_i$ into 0~K local dipole moments $P_{0i}$ via their Born effective charge tensors ${\mathbf Z}^*_i$
\footnote[2]{Following approach of Ref. 26, we used approximate $Z^*$ values of 3.6, 2.6, 2.6, 7.4, 4.0, 12.0, 6.4 and 6.4 for Pb, Mg, Zn, Nb, Sc, W, Ti and Zr respectively. Oxygen $Z*$ were $\approx$ -5.0 and $\approx$ -2.5 for the parallel and perpendicular elements of the tensor respectively, varied slightly to satisfy the electrostatic sum rule for each composition.}.  
This leads to

\begin{eqnarray}
  T_{\epsilon,\rm max} =  \gamma \overline{|P_{0i}|}^2 =  \gamma \left( \sum_i \frac{|\mathbf{P}_{0i}|}{N}\right)^2=\gamma \left(\sum_i  \frac{|{\mathbf Z}^*_i \cdot {\mathbf u}_i|}{NV_0}  \right)^2  {\rm or} 
\end{eqnarray}

\begin{eqnarray}
 T_{\epsilon,\rm max} =  \gamma P_0^2 = \gamma \left( \sum_i \frac{\mathbf{P}_{0i}}{N}\right)^2=\gamma \left(\sum_i  \frac{{\mathbf Z}^*_i \cdot {\mathbf u}_i}{NV_0}  \right)^2\ 
\end{eqnarray}

\noindent  where  $\gamma$ is a constant, $N$ is the number of primitive unit cells and $V_0$ is the volume of a primitive unit cell  and $\overline{|P_{0i}|}$ and  $P_0$ are the 0~K  average magnitude of the local dipole moment and the overall polarization respectively.   Figure 3 shows the correlation between the $T_{\epsilon,\rm max}$ predicted by Eqns.\ 1-3 and experimental $T_{\epsilon,\rm max}$ data~\cite{Landolt}.   The clustering of the data around the $y$=$x$ line indicates that Eqns.\ 1-3 accurately and quantitatively capture the trends in $T_{\epsilon,\rm max}$ among these fairly different systems.   Comparing the data for PMN-PT and PZN-PT systems in Table I, we see that the  higher $T_{\epsilon,\rm max}$ for PZN-PT system is directly related to larger $P_0$  and $\overline{|P_{0i}|}$ values due to off-centering behavior on the B$^{\rm 2+}$ site.

The correlations presented in Figure 3 can be explained based on fourth-order Landau theory for ferroelectric crystals.    The equilibrium polarization is given by

\begin{eqnarray}
dG/dP =0=2\alpha(T-T_0)P+4\beta P^3
\end{eqnarray}

\noindent  where $\alpha$ and $\beta$ are constants, $T_0$ is the Curie-Weiss temperature at which the dielectric constant $\epsilon$ diverges and $P$ is polarization. 
Substituting $T$=0 used in our DFT calculations and solving for $T_0$, we obtain $ T_0 =  2\beta P_0^2/\alpha$,
thus relating the $T_0$ to the square of computed 0~K polarization.  The high quality of the fit in Figure 2 means that for the Pb-based perovskites examined here the ratio $\beta$/$\alpha$  is nearly constant.

The energy difference between ferroelectric and paraelectric phases (the ferroelectric instability $\Delta E_{\rm FE}$) is commonly thought to correlate with $T_0$.  However, $\Delta E_{\rm FE}$ does not correlate with $T_0$ for Pb-based perovskites such as PZT~\cite{Fornari01p092101}, PSW-PT~\cite{Juhas04p214101} and (Cd,Pb)TiO$_3$~\cite{Halilov03p3443,Suarez-Sandoval03p3215} (CPT).  The constancy of $\beta$/$\alpha$ helps explain this observation. In fourth-order Landau theory $\Delta E_{\rm FE}$ at $T=0$ is given by 
$\Delta E_{\rm FE}=\alpha^2 T_0^2/4\beta$,
  with $T_0^2$=$(4\beta/\alpha^2)\Delta E_{\rm FE}$. The $\beta$ parameter scales the leading order repulsive term and corresponds to the repulsive potential felt by the off-centering Pb and B-cations.  This potential varies from system to system based on different sizes of the B-cations~\cite{Grinberg02p909,Grinberg04p144118}.  Since the ratio of $\beta/\alpha$ is constant for a wide variety of ferroelectric solutions as shown above, the value of $\beta/\alpha^2$ varies. This makes $\Delta E_{\rm FE}$ a poor predictor of $T_0$. 

The nearly constant dependence of $T_{\epsilon,\rm max}$ on $P_0^2$ and $\overline{|P_{0i}|}$  can be interpreted as follows.  Experimental results have shown that ferroelectric to paraelectric phase transitions in Pb based perovskites such as PMN and PZT are of order-disorder character~\cite{Egami03p48}, with Pb displacements at $T_0$ of up to 70$\%$ of the low temperature value.  The parameter governing the temperature of such an order-disorder transition  is the coupling strength between the Pb and B-cation displacements in neighboring unit cells  (corresponding to the $J$ parameter in spin models).  The coupling is due to electrostatic effects favoring dipole alignment and covalent chemical bonding effects disfavoring under- and over-bonded oxygen atoms.   Dipole alignment strengths depend only on polarization strengths and gives rise  to the overall dependence of $T_0$ on $P_0^2$ or $\overline{|P_{0i}|}^2$, with the variation in through-oxygen coupling accounting for the differences between experimental and predicted $T_{\epsilon,\rm max}$ values in Figure 3.   While predicted $T_{\epsilon,\rm max}$ values are of the same quality for Eqn.\ 1 and Eqn.\ 2, the correlation between  $P_0$ and $T_{\epsilon,\rm max}$ is somewhat worse.  This suggests that the freezing in of large distortions enhances stability of ferroelectricity, giving rise to higher $T_{\epsilon,\rm max}$.  Even though some of the distortions frozen in are not parallel, they still contribute to the coupling of dipoles in neighboring unit cells due to through-oxygen interactions.

In summary, we have examined structure-property correlations for PMN-PT and PZN-PT solid solutions. We find that the local direction of Pb displacements in these systems is governed by an interplay between the electrostatic dipole-dipole alignment interactions and Pb-B repulsive interactions, as was previously found for PZT.  
We find that Zn ions off-center; this reduces Pb-B repulsions resulting in smaller Ti content at MPB and allowing larger displacements by the Pb, Nb and Ti ions. 
A strong relationship exists between the cation displacements or $P_0$ and $T_{\epsilon,\rm max}$, which can be explained in the Landau theory framework by the constant ratio of $\alpha$ and $\beta$ parameters and is a natural consequence of the order-disorder nature of phase transitions in Pb-based ferroelectrics.  The link between $\alpha$ and $\beta$ also explains the opposite trends in $\Delta E_{\rm FE}$ and $T_{\rm c}$ found in the PZT, PSW-PT and CPT systems.   Fundamentally,  we have shown that behavior of individual ions in heterovalent perovskite ferroelectric solid solutions can be directly linked  to changes in local structure and macroscopic collective properties.

This work was supported by the Office of Naval Research, under grant number N-000014-00-1-0372 and through the Center for Piezoelectrics by Design. We also acknowledge the support of the National Science Foundation, through the MRSEC program, grant No. DMR00-79909.  AMR acknowledges the support of the Camille and Henry Dreyfus Foundation.   Computational support was provided by the Center for Piezoelectrics by Design, the  DoD HPCMO and DURIP.  We would also like to thank Peter K. Davies for stimulating discussions.

\bibliography{apssamp}% Produces the bibliography via BibTeX.

%\noindent Correspondence and requests for materials should be addressed to I.G. (email: ilya2@sas.upenn.edu)

\clearpage
\begin{table}[!t]
\caption{Results of DFT calculations for PMN-PT and PZN-PT systems.  Pb displacement magnitudes in PMN predicted by DFT are in agreement with recent experimental data~\cite{Egami03p48} and $P$ magnitudes for PMN and PT are in good agreement with previous theoretical calculations~\cite{Prosandeev04preprint}. Predicted cation displacements from center of oxygen cage in \AA, Pb displacement angle scatter $\theta_{\rm Pb}$ in $^\circ$, average and local polarization in C/m$^2$ and experimental $T_{\epsilon,\rm max}$~\cite{Landolt} in K.}

\begin{tabular}{lcccccc}

&\multicolumn{1}{c}{Pb}
&\multicolumn{1}{c}{B$^{2+}$}
&\multicolumn{1}{c}{Nb/Ti}
&\multicolumn{1}{c}{$\theta_{\rm Pb}$}
&\multicolumn{1}{c}{$P_0$}
&\multicolumn{1}{c}{$T_{\epsilon,\rm max}$}\\ 

&\multicolumn{1}{c}{disp}
&\multicolumn{1}{c}{disp}
&\multicolumn{1}{c}{disp}
&\multicolumn{1}{c}{}
&\multicolumn{1}{c}{avg,loc}
&\multicolumn{1}{c}{exp.}\\ 

\hline
PMN 	   &	0.398&	0.056&	0.168&	65&	0.38, 0.67	& 276\\	
PMN-0.25PT &	0.389&	0.080&	0.181&	33&	0.55,0.68	& 397\\	
PMN-0.625PT &	0.387&	0.099&	0.216&	28&	0.66, 0.74	& 583\\	
PT	   &	0.440&	     &	0.280&	0&	0.88,0.88	& 765\\	

 	   &	& & &	&		&\\
	
PZN 	   &	0.444&	0.148&	0.189&	67&     0.43,0.73       & 424\\	
PZN-0.25PT &	0.461&	0.258&	0.205&	35&	0.66,0.79	& 547\\  
PZN-0.625PT &	0.424&	0.270&	0.237&	27&	0.74,0.80	& 643\\	  
PT	   &	0.440&	     &	0.280&  0&	0.88,0.88	& 765\\

\end{tabular}
\end{table}

\clearpage
\begin{figure}
\includegraphics[width=4.0in]{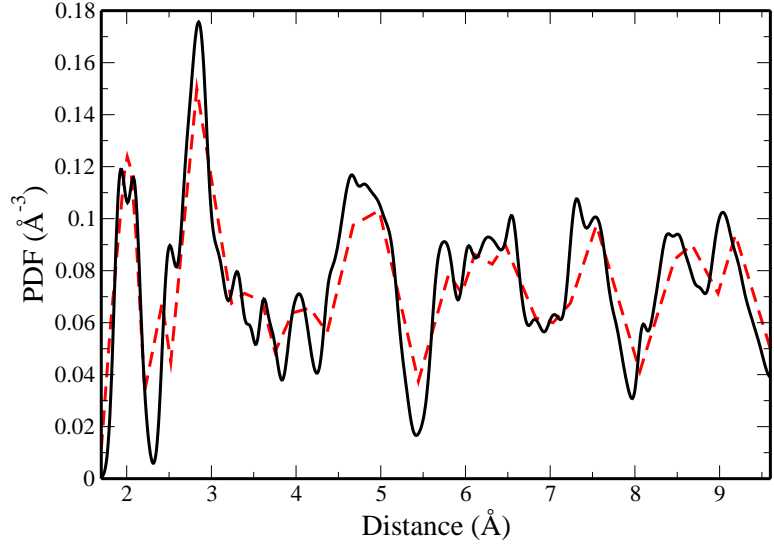}
\caption{{Comparison of PDFs obtained from relaxed structure of 60-atom supercell of PMN (solid) and by neutron-scattering (dashed)~\cite{Egami98p1}.  Similar agreement is obtained for other systems~\cite{Grinberg04p144118,Juhas04p214101}.}}

\end{figure}

\begin{figure}
\includegraphics[width=4.0in]{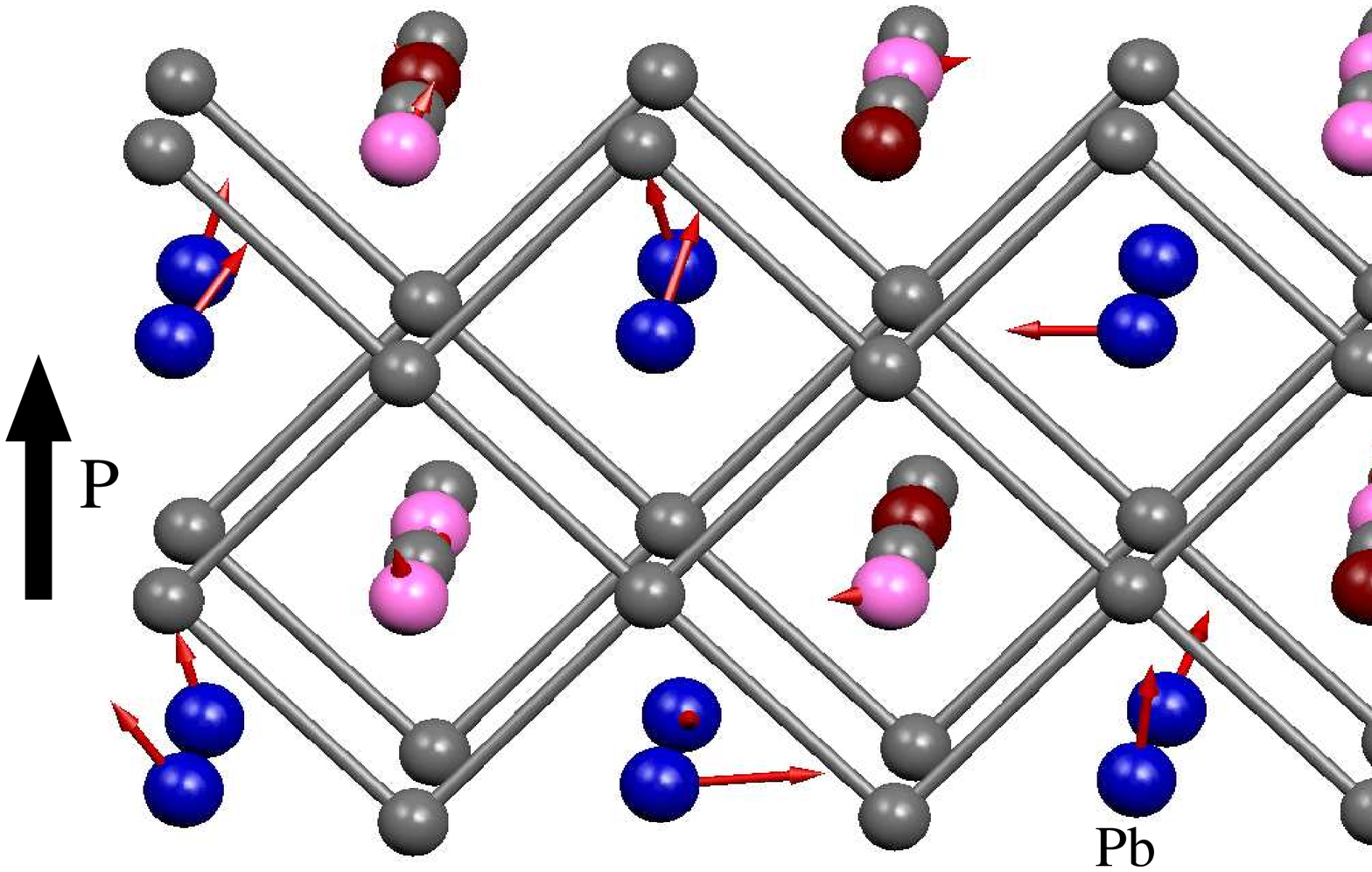}
\caption{{ Relaxed structure of 60-atom supercell of PMN obtained by DFT calculations. Cation displacements from high-symmetry positions are shown by arrows scaled up by a factor of 4.  Pb atoms displace away from Nb-rich faces. Blue, pink, red and gray represent Pb, Nb, Mg and O atoms respectively. }}
\end{figure}

%\begin{figure}
%\includegraphics[width=4.0in]{partial.eps}
%\caption{{ Pb-B-cation partial PDFs obtained from DFT calculations for PMN and PZN.   Pb-Mg and Pb-Zn peaks are located at shorter distances than the Pb-Nb peaks due to the stronger through-oxygen Pb-Nb repulsion.}}
%\end{figure}

\begin{figure}
\includegraphics[width=4.0in]{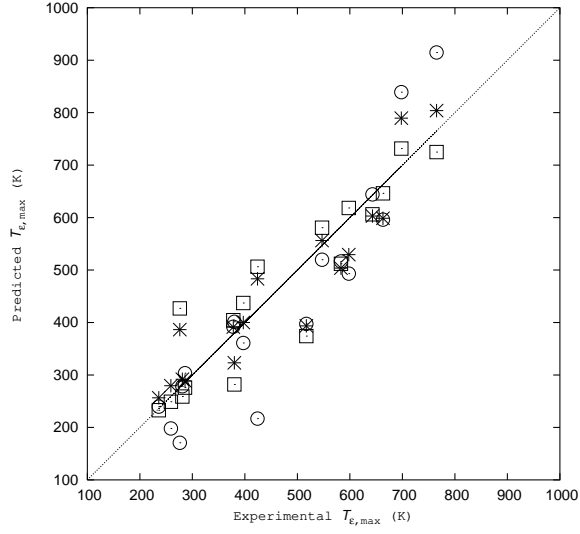}
\caption{{ Correlations between experimental $T_{\epsilon,\rm max}$ and predicted $T_{\epsilon,\rm max}$ for PZN, PZN-PT, PMN, PMN-PT, PSW, PSW-PT, PSW-PZ, PSN and PZT systems. Values for predicted $T_{\epsilon,\rm max}$ are obtained by Eqn. 1 (squares) with $a$=1739 and $b$= 5961, Eqn. 2 using local polarization with $\gamma$=942 (stars) and Eqn. 2 using overall polarization with $\gamma$=1189 (circles).  }}
\end{figure}

\appendix

%\noindent {\bf References} 

\end{document}